\documentclass{article}

\usepackage{arxiv}

\usepackage[utf8]{inputenc} 
\usepackage[T1]{fontenc}    
\usepackage{hyperref}       
\usepackage{url}            
\usepackage{booktabs}       
\usepackage{amsfonts}       
\usepackage{nicefrac}       
\usepackage{microtype}      
\usepackage{lipsum}		
\usepackage{graphicx}
\usepackage{natbib}
\usepackage{doi}

\title{Ocean-DC: An analysis ready data cube framework for
environmental and climate change monitoring over the port areas}

\author{ \href{https://orcid.org/0000-0003-0448-9459}{\includegraphics[scale=0.06]{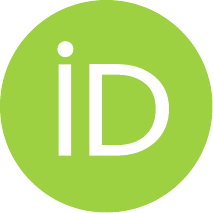}\hspace{1mm}Ioannis Kavouras}\\
	School of Rural, Surveying and Geoinformatics Engineering\\
	National Technical University of Athens\\
	9, Iroon Polytechniou st, 15772, Zografou \\
	\texttt{ikavouras@mai.ntua.gr} \\
	\And
	\href{https://orcid.org/0000-0003-4491-5854}{\includegraphics[scale=0.06]{orcid.pdf}\hspace{1mm}Ioannis Rallis}\\
	School of Rural, Surveying and Geoinformatics Engineering\\
	National Technical University of Athens\\
	9, Iroon Polytechniou st, 15772, Zografou \\
	\texttt{irallis@central.ntua.gr} \\
    \And
	\href{https://orcid.org/0000-0002-4064-8990}{\includegraphics[scale=0.06]{orcid.pdf}\hspace{1mm}Nikolaos Doulamis}\\
	School of Rural, Surveying and Geoinformatics Engineering\\
	National Technical University of Athens\\
	9, Iroon Polytechniou st, 15772, Zografou \\
	\texttt{ndoulam@cs.ntua.gr} \\
    \And
	\href{https://orcid.org/0000-0002-0612-5889}{\includegraphics[scale=0.06]{orcid.pdf}\hspace{1mm}Anastasios Doulamis}\\
	School of Rural, Surveying and Geoinformatics Engineering\\
	National Technical University of Athens\\
	9, Iroon Polytechniou st, 15772, Zografou \\
	\texttt{adoulam@cs.ntua.gr} \\	
}

\renewcommand{\shorttitle}{Ocean-DC: An analysis ready data cube framework for
environmental and climate change monitoring over the port areas}

\begin{document}
\maketitle

\begin{abstract}
The environmental hazards and climate change effects causes serious problems in land and coastal areas.
A solution to this problem can be the periodic monitoring over critical areas, like coastal region with
heavy industrial activity (i.e., ship-buildings) or areas where a disaster (i.e., oil-spill) has occurred.
Today there are several Earth and non-Earth Observation data available from several data providers. These
data are huge in size and usually it is needed to combine several data from multiple sources (i.e., data
with format differences) for a more effective evaluation. For addressing these issues, this work proposes
the Ocean-DC framework as a solution in data harmonization and homogenization. A strong advantage of this
Data Cube implementation is the generation of a single NetCDF product that contains Earth Observation data 
of several data types (i.e., Landsat-8 and Sentinel-2). To evaluate the effectiveness and efficiency of the
Ocean-DC implementation, it is examined a case study of an oil-spill in Saronic gulf in September of 2017.
The generated 4D Data Cube considers both Landsat-{8,9} and Sentinel-2 products for a time-series analysis,
before, during, and after the oil-spill event. The Ocean-DC framework successfully generated a NetCDF product,
containing all the necessary remote sensing products for monitoring the oil-spill disaster in the Saronic
gulf.
\end{abstract}


\section{Introduction}
\label{sec::Introduction}

The modern urban environments have been developed with the purpose of interconnecting people, 
infrastructures, activities and several other resources, making them susceptible to environmental
disasters and climate change effects \cite{mi2019cities, leal2019assessing, olabi2022renewable}.
Daily industrial activities, especially near the port areas, have a negative impact on the
local environment and climate endangering both the human health and well-being, as well as the
local nature species. 

All the human activities result in the following issues: (a) lack of urban greenness \cite{temenos2022novel};
(b) flooding \cite{pour2020low}; (c) non-healthy atmospheric quality \cite{ravindra2019generalized};
(d) increment of greenhouse gases emissions \cite{shen2020micro}; and (e) the phenomenon of heat island effect 
\cite{parker2010urban}. In most of these cases, the impact can be evaluated by continuously monitoring the
activities over the area of interest using Earth Observation (EO) data provided by space-borne technologies 
(i.e., satellites) and aerial or field climate measurements. In other cases, the impact can be instantaneous, 
causing a huge disaster over a radius, like in an oil-spill situation.

Today, there is available a large number of EO data and it expands each week \cite{gomes2020overview}. Moreover,
the available data are produced from different data providers (i.e., Copernicus, USGS, ESA, NASA, etc.), thus there
are differences in distribution format, spatial resolution, data collection, etc. In some cases, however, two missions
from different provides (i.e., Landsat-8 and Sentinel-2) have similarities, thus such data can be used in combination
for thickening a time-series dataset. For a successful data harmonization and homogenization a Data Cube (DC) implementation
can be used \cite{kopp2019achieving}.

\begin{figure*}[!htb]
  \centering
  \includegraphics[width=0.75\linewidth]{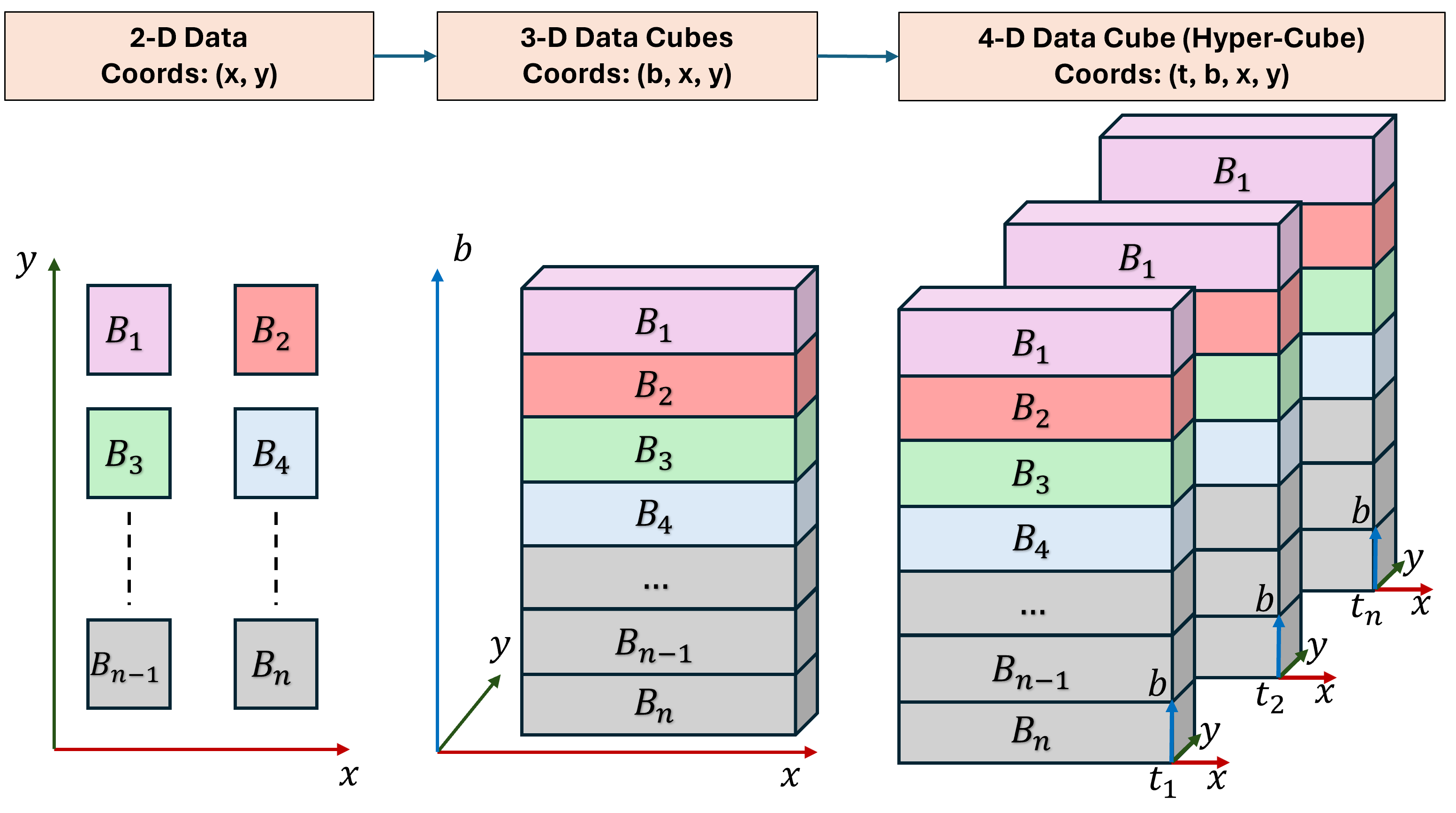}
  \caption{An example of a Data Cube's dimension expanding. \label{fig::Intro::DataCube}}
\end{figure*}

A DC \cite{baumann2019datacubes, harinarayan1996implementing} is a multidimensional array containing all the necessary
information (i.e., data, metadata, coordinates, etc.) describing a phenomenon. To better understand a basic Earth Observation
DC framework, let us denote as 2D array ($B_i(x, y)$) corresponding an spectral band (i.e., image) of a satellite product. Thus, 
each coordinate set (x, y) corresponds to a radiometric reflectance value for this band. However, each satellite retrieves
data on several wavelengths, which means that they generate several images for different wavelengths of the spectrum (i.e.,
Coastal Aerosol, Blue, Green, Red, Near Infrared, etc.). The combination of these images can generate several remote sensing
products \cite{zhang2010principles, wu2009scale} for indicating a phenomenon. Usually, the raw data are captured in different
spatial resolutions (i.e., 10m/px, 15m/px, 30m/px, 60m/px, etc), thus any combinations and/or calculations can not be achieved in
the original sate of data. For addressing this issue, let us now denote a 3D array ($DC_i(b, x, y)$) for storing all the spectral
bands $N$ ($DC_i(b, x, y) = [B_1(x, y), B_2(x, y), ..., B_N(x, y)]$) in a common spatial resolution format and coordination reference 
system. Finally, the monitoring of a phenomenon introduces the parameter of time ($t$). Thus, for effectively monitoring a periodic
phenomenon (i.e., climate) over an are and detecting abnormalities it is necessary to generate a DC containing data from various sources.
Thus, a final DC implementation can be a 4D-DC (Hypercube) defined as $HDC_i(t, b, x, y) = [DC_1(b, x, y), DC_2(b, x, y),... DC_N(b, x, y)]$,
containing $N$ 3D-DC of different time periods (i.e., before and after an event). Figure \ref{fig::Intro::DataCube} illustrates the DC dimension
expanding as described in this paragraph.

For the successful implementation of a 4D-DC it is necessary to harmonize and homogenize all of the available 3D-DCs. If the original data are 
provided from the same data provider, then the implementation can follow the same processing, however additional processes are necessary if the 
data are provided from more than one sources. For addressing this challenge, in this work, we propose the Oceanids Data Cubes (Ocean-DCs) as a 
novel approach for continuously and quickly monitoring an area of interest using several kinds of data from the same or different data sources. 
Moreover, the Ocean-DC provides data harmonization and homogenization by applying the common template for Landsat-8 and Sentinel-2 data products, 
as well as it offers a complete ready to use analysis products by calculating the most common used remote sensing indices. To be more precise, 
Ocean-DCs' products can be used as is in a decision support system platform alongside in synergy with other data of the platform for indicating 
environmental and climate change hazards in many applications, emphasizing to port, marine and coastal applications (i.e., monitoring, forecasting, 
etc.).

For demonstrating the Ocean-DC implementation a monitoring case study is provided in this manuscript. For the case study it is selected the oil-spill
disaster that took place on the 10th of September 2017 in the Saronikos gulf (Greece) causing huge environmental disaster to the surrounded shores and
sea/water area. This case study aims to provide an application example of the proposed Ocean-DC for periodically monitoring a phenomenon based on 
time-series analysis. However, Ocean-DC products can be used differently depended on the application and the related phenomenon.

The rest of this paper is organized as follow: (a) Section \ref{sec::RelatedWork} presents the related literature
review; (b) Section \ref{sec::OceanDCframework} describes the proposed Ocean-DC framework; (c) Section \ref{sec::CaseStudyAnalysis}
provides a brief demonstration of environmental monitoring based on an Ocean-DC product; (d) Section \ref{sec::Discussion} discuss the outcomes
of this work and (e) Section \ref{sec::Conclusions} summarizes this work.

\section{Related Work}
\label{sec::RelatedWork}

A Data Cube is a multidimensional array capable or a data structure that allows the efficient analysis of big data and complex data
structures along the multiple dimensions. A DC framework is mostly used for applications regarding big data storage and online analytical
processing (OLAP) \cite{sarawagi1998discovery} based on complex queries \cite{gray1997data} and analysis \cite{francia2022enhancing}. Major
functionalities of a DC framework are: (a) multidimensional representation \cite{leprince2021data}; (b) cube structure for big data 
visualizations \cite{song2018multidimensional}; (c) data aggregation \cite{giuliani2020monitoring}; (d) fast query performance
\cite{miranda2017topkube}; (e) OLAP operations \cite{tardio2020new}; and (d) smart big data applications \cite{majeed2021big}.

The current DC’s implementations are emphasizing on organizing the data on multiple dimensions \cite{han2005stream} like time, geography, 
product, or customer. Each dimension represents a different aspect of the data. Thus, DCs structure is often visualized as cube with multiple 
dimensions. Each cell in the cube’s representation is a unique combination of dimensional values. For example, in a remote sensing DC an 
indicative combination can be the (time x spectral bands x geographical coordinates). Moreover, in other DC examples, like climatic and 
geographical DC, the data can be aggregated using traditional aggregation functions like mean, count, average, minimum, and maximum.

However, it is important to note that the numerical values alone are not enough to describe a phenomenon. Another strong characteristic of 
the DC’s architecture is the inclusion of metadata information that describes the numerical values of the data \cite{boulil2015conceptual}. 
Such information includes the metric system of the data, the origin of the data, geo-spatial information, etc. This information can be used 
for achieving fast query performance, which is one of the main advantages of the utilization of DCs, especially when dealing with large volumes 
of data (i.e., addressing big data storage performance issues).

Several Data Cube implementations have been suggested the recent years. An indicative
example can be the work of Ferreira et al. \cite{Ferreira2020earth}, where they proposed
an effectual way for generating Earth Observation Data Cubes (EODCs) over the Brazil region.
They additionally are exploiting the benefits of the Open Data Cubes (ODC) platform, contributing
by conducting a analysis-ready data and multidimensional data cubes from remote sensing images 
to effectively mapping the land use and land cover based on employing remote sensing imagery
time series analysis and machine learning techniques.

The Australian Geoscience Data Cubes (ADGCs) is one of the first data cubes implementation,
proposed by Lewis et al. \cite{Lewis2017big} and manages the challenges of the Big Data, such
us the storage volume, velocity, and polymorphism or variety, limiting the earth observation data
usefulness. Some of the application, where the ADGCs have been applied are: (a) water observations;
(b) coastal erosion; (c) agriculture monitoring; (d) forest cover changes; and (e) biodiversity.

In the work of Poussin et al. \cite{Poussin2019snow} the advantages and limitations of the Earth Observation
Swiss Data Cubes (EOSDCs) are exploited by monitoring the snow cover change in Gran Paradiso national Park, Alps.
EOSDCs architecture is based on the Open Data Cubes software \cite{chatenoux2021swiss}, which is a geospatial
data management and analysis software based on open-source technologies for processing remote sensing data by
providing a framework for accessing, storing, managing, and analyzing grid-aligned satellite Earth Observation
data collections in huge quantities. 

Other works utilizing the EOSDCs implementations are proposed by Giuliani et al. 
\cite{giuliani2020monitoring, giuliani2020data}. In the first work \cite{giuliani2020monitoring} they propose a
methodology for generating accurate and consistent land degradation products based on remote sensing imagery datasets.
They uses a data cube and several sun-indicators for assessing land productivity and cover, as well as soil organic
carbon changes for evaluating land degradation. The second work \cite{giuliani2020data} suggests the usage of a data
cube on demand (DCoD), which is a script-based tool, for automating the generation of data cubes under predetermined by
the user requirements. Such requirements can be the area of interest, a time period, the type of the sensor, or the
satellite product family. Their implementation has been successfully applied to two regions (Bolivia and Democratic Republic
of Congo) for environmental monitoring.

Finally, Temenos et al. \cite{temenos2024c2a} propose a context-aware adaptive data cube (C2A-DC) framework, which utilize
Earth Observation data for environmental monitoring and mitigating the climate change effects. Their proposed methodology
combines the data cubes formation with calculation of remote sensing operations, including deep and machine learning algorithms for 
data classification and harmonization. Thus, their outcome is a context-aware adaptive framework, which successfully allows the
environmental monitoring and climate change effects based on user predetermined preferences.

\subsection{Contribution}

Considering the current literature is is noted that several DC implementations have been proposed for different applications over
specific regions. However, these implementations are focusing only on the harmonization and homogenization of data provided from the
same source, even if they provide DCs of several data sources. As an example, the current DC implementations can provide different data
like Landsat-8, Sentinel-2, Sentinel-3, etc., however each cube contains data only from one source. In addition, the current
DC implementations provide additional products (i.e., NDVI, NDWI, etc.) on demand based on some query selection.

This work approach the problem of handling the big data from a different scope. The proposed Ocean-DC implementation provide data
harmonization and homogenization in a 4D-DC scenario, including data from several sources with commonalities in the same cube. For
example, in the case study of this manuscript are used Landsat-8 and Sentinel-2 data for the time series analysis, that are included 
in the same 4D-DC. In addition, the Ocean-DC pre-calculates the most common remote sensing products (e.g., NDVI, NDWI, Land Surface 
Temperature, etc.) providing to the end-users a ready to use product for addressing several environmental and climate change challenges.

\section{Proposed Ocean-DC framework}
\label{sec::OceanDCframework}

\begin{figure*}[!htb]
  \centering
  \includegraphics[width=\linewidth]{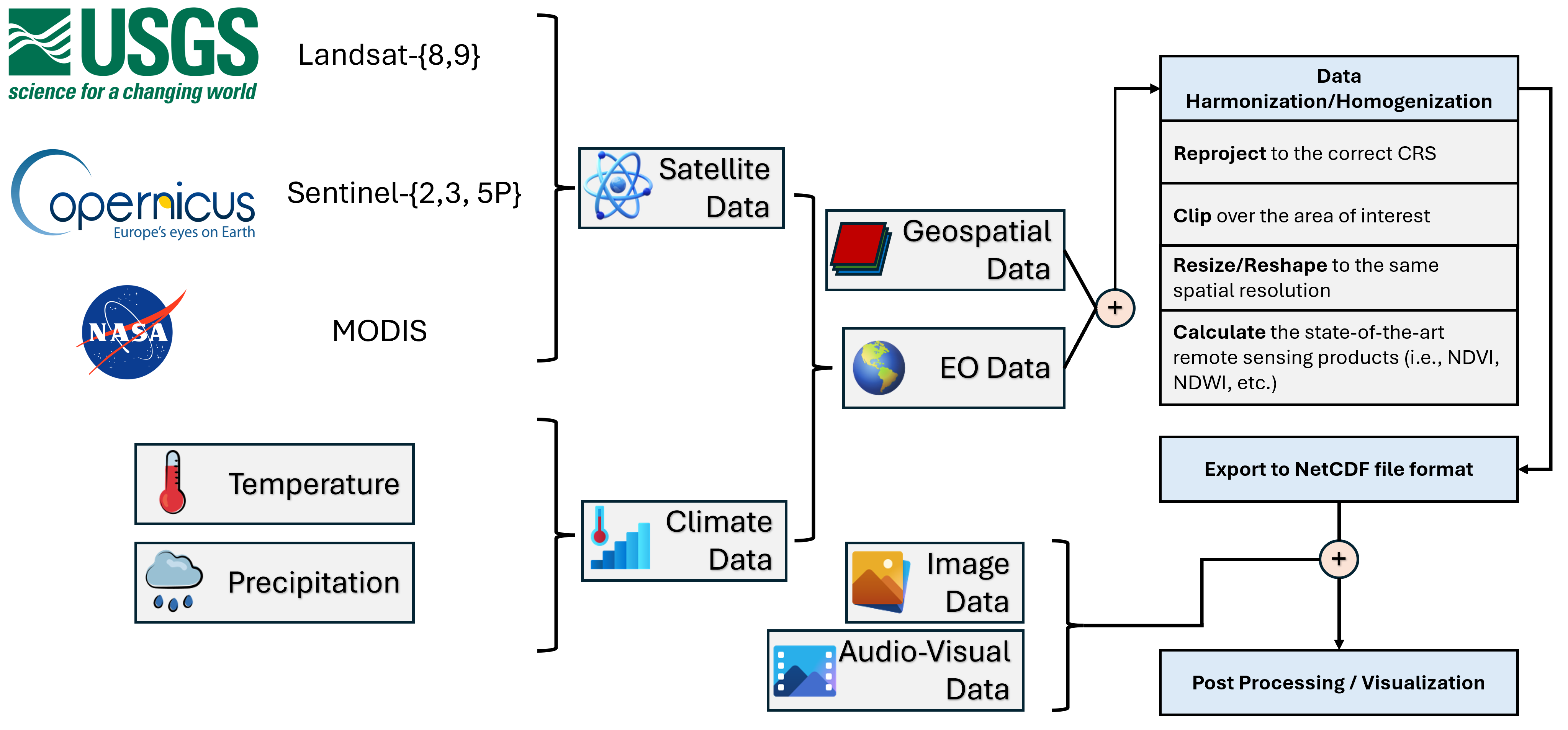}
  \caption{The workflow architecture of the Ocean-DC implementation.}
  \label{fig::PropMeth::Architecture}
\end{figure*}

The main objective of the Ocean-DC implementation is to provide harmonization and homogenization between the several available
remote sensing and climate data products (i.e., Earth Observation [EO] products), that can be retrieved from multiple data providers 
(i.e., USGS, Copernicus, NASA). In addition, non-EO data can be used, like the boundaries of the area of interest
in the processes. All these data are collected in the first step of the whole framework architecture, as illustrated in Figure
\ref{fig::PropMeth::Architecture}.

To be more precise, the Data Collection considers several EO and non-EO data including geolocated raster remote sensing
and climate products and the non-EO products are usually vector type. Currently, the Ocean-DC implementation supports
products from USGS \cite{EarthExplorer} (i.e., Landsat-{8,9}), Copernicus Services \cite{CopernicusHub} (i.e., Sentinel-{2,3,5P} and
CERRA climate data) and NASA \cite{MODIS} (i.e., MODIS) EO data. The supported non-EO data are polygon shapefile (SHP) geometries. 
These data are fed to the next step (i.e., Data Harmonization/Homogenization) for generating the Ocean-DCs.

The Data Harmonization/Homogenization step is the core tool \cite{OceanDC-framework} of the workflow architecture. In this step,
the EO products are reprojected to a common Coordinate Reference System (CRS) to get the same coordinates per pixel. Usually, this
coordinate system is the same of the SHP geometry, but the user can select any CRS or EPSG. This step is necessary because the
EO data are in different CRS. For instance Landsat-8 and Sentinel-2 raw data are provided in different CRS. Thus, this step is
necessary before applying any other process.

After the data reprojection, the framework clips the products to a box geometry based on the user input (i.e., SHP geometry).
The box geometry is calculated by the minimum and maximum (x,y) coordinates of this SHP geometry. Image clipping is a critical
step, because it significantly minimizes the size of the original images and as a natural result the size of the final Ocean-DC
product. After successfully clipping the raster EO products, they are in the same coordinate reference system and visualize the
area of interest.

The next following two steps are the image resizing/resampling and the calculation of the most used remote sensing product, in the
case of Landsat-8 and Sentinel-2 EO products. The image resizing/resampling process corrects the spatial resolution of the images 
and ensures that all images have the same width and height pixels. Note that it is not added new information to the original data.
This means that an image with spatial resolution 60m/px will be resampled to 10m/px by splitting each pixel to a 6x6 grid with the
same value (i.e., pseudo-resolution). This step increases the size of the final product, but it is very important, because the
data cube cannot be constructed with different image resolutions. In addition, it is impossible to perform raster calculation with
images in different resolutions. Thus, the image resizing/resampling is a mandatory step to proceed in the next steps.

Finally, the calculation of the most common remote sensing products is performed when it is necessary. To be more precise, the
remote sensing indices are calculated when specific band information is known (i.e., optical bands, near-infrared, SWIR, etc.).
Currently, this step is applied for Landsat-{8,9} and Sentinel-2 products. Table \ref{tab::BandSummary} summarizes the satellite 
calculated products, which are containing in an Ocean-DC implementation. The processing of the other products is terminated in the
image resizing/resampling step.

The implementation of the Ocean-DC is performed using Python and the state-of-the-art and open-source DC handling library 
Xarray \cite{Xarray}. This library is empowered with the ready to use methods of rasterio \cite{Rasterio}, which is another
python library for handling EO data. The final outcome of the data harmonization/homogenization framework is a NetCDF file,
which is a state-of-the-art format for geodata and is supported by several GIS software like QGIS \cite{QGIS}.

Finally, the Ocean-DC products can be combined with other non-EO data like images or drone footage to produce advanced 
cartographic products or monitor several phenomena. Some examples of post-processing applications can be the monitoring
of a disaster (i.e., flooding, forest fire, oil-spill, etc.) or the generation of thematic maps that describes a phenomenon.
Other application could be the monitoring of air quality or climate factors (i.e., temperature, precipitation).

\begin{table}[!ht]
    \centering
    \caption{Summary of the products, where containing in an Ocean-DC file. \label{tab::BandSummary}}
    \begin{tabular}{c l | c l }
    \hline
        \textbf{BAND ID} & \textbf{Description} & \textbf{BAND ID} & \textbf{Description} \\ \hline
        1 & COASTAL AEROSOL & 23 & LSWI-1 \\
        2 & BLUE & 24 & LSWI-2 \\
        3 & GREEN & 25 & ARVI \\
        4 & RED & 26 & MSAVI2 \\
        5 & NIR & 27 & MTVI2 \\
        6 & SWIR-1 & 28 & VARI \\
        7 & SWIR-2 & 29 & TGI \\
        8 & CIRRUS & 30 & LST-1 \\
        9 & VRE-1 & 31 & LST-2 \\
        10 & VRE-2 & 32 & LST-CELSIUS \\
        11 & VRE-3 & 33 & LST-FHRENHEIT \\
        12 & VRE-4 & 34 & VCI \\
        13 & WATER VAROUR & 35 & MNDWI-1 \\
        14 & PANCHROMATICS & 36 & MNDW-2 \\
        15 & TIRS-1 & 37 & WRI-1 \\
        16 & TIRS-2 & 38 & WRI-2 \\
        17 & NDVI & 39 & NDTI \\
        18 & NDWI & 40 & AWEI \\
        19 & NDVI  (VRE-1) & 41 & OSI \\
        20 & NDVI (VRE-2) & 42 & NBR-1 \\
        21 & NDVI (VRE-3) & 43 & NBR-2 \\
        22 & NDVI (VRE-4) \\ \hline
    \end{tabular}
\end{table}

\section{Case Study Analysis of the oil-spill in Saronikos gulf at Septempber 2017}
\label{sec::CaseStudyAnalysis}
For the demonstration of the Ocean-DC framework is used the oil-spill disaster that took place in Saronic gulf in 
10th of September 2017. This disaster has been also investigated in the work of Kolokoussis and Karathanassi \cite{kolokoussis2018oil}
where they use the NDWI among other indices to visually enhance the critical areas of the oil-spill. In a similar way, this work
investigates the oil-spill disaster emphasizing mostly on the possibilities that Ocean-DC framework provides to the end-users.
For this reason, in this analysis it is used a dataset composed of Landsat-8 and Sentinel-2 products in a time-period before the
oil-spill (2016 to 9th of September 2017), some days and weeks after the the oil-spill (10th of September to October 2017) and
recent images (2023-2024). Thus, the implemented Ocean-DC product is consisted of 16 satellite imagery products (5 Landsat-8 and 11 
Sentinel-2). For visualizing the oil-spill effectively it used the NDWI, WRI-2, and OSI \cite{Zakzouk2024novel} indices for monitoring 
the water quality of Saronic gulf during the aforementioned periods.

\begin{figure*}[!htb]
  \centering
  \includegraphics[width=0.85\linewidth]{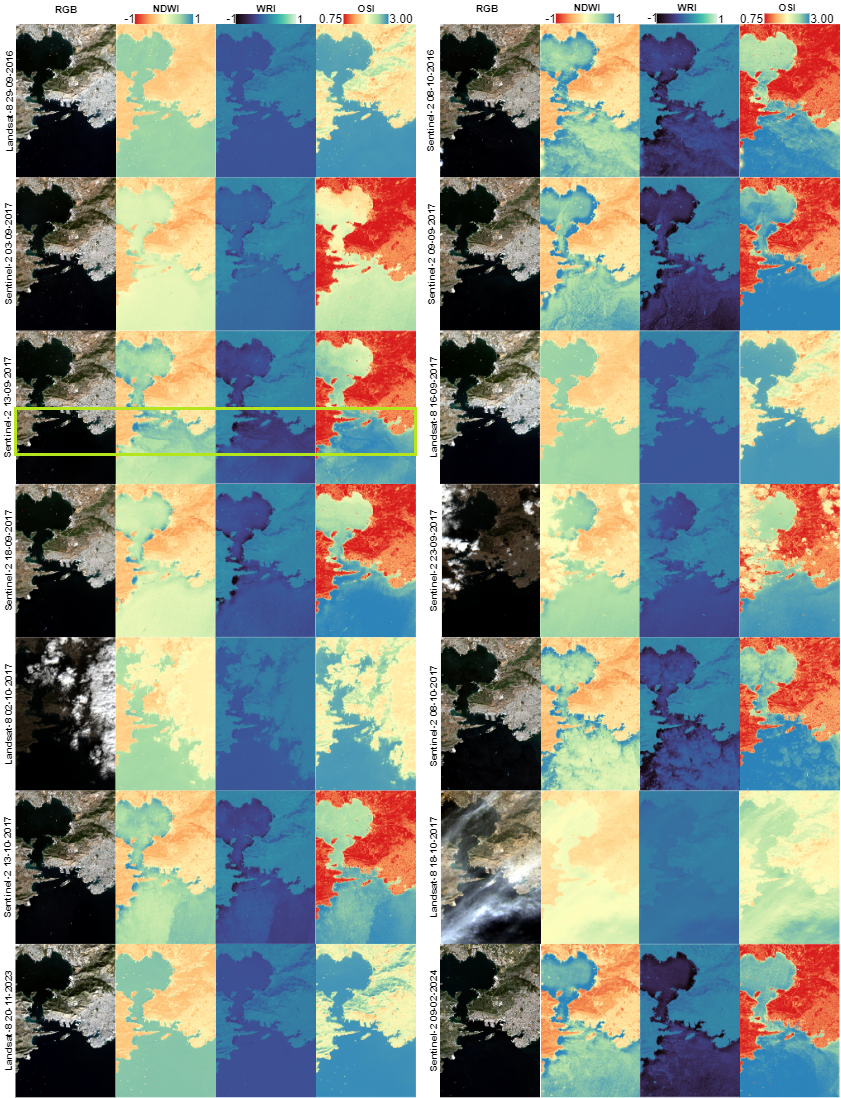}
  \caption{Time series monitoring of the water quality of Saronikos gulf before and after the oil-spill using the RGB, NDWI, WRI, and OSI 
  products of the Ocean-DC implementation file. The green box indicates the area of the oil-spill. \label{fig::CaseStudyAnalysis::Results}}
\end{figure*}

The results of the generated Ocean-DC product are illustrated in Figure \ref{fig::CaseStudyAnalysis::Results}. The visualization of the
generated product performed inside QGIS software and considers the natural (RGB) color composition, the NDWI, WRI, and OSI indices illustrated 
using pseudo-palette colors. The NDWI product is calculated using the Equation \ref{eq::NDWI} and the output's image values are in range 
$[-1.0, 1.0]$ as follows: $[-1.0, -0.3) \rightarrow$ Drought, Non-Aqueous Surfaces; $[-0.3, 0.0) \rightarrow$ Moderate drought,
non-aqueous surfaces; $[0.0, 0.2) \rightarrow$ Flooding, humidity; $[0.2, 1.0] \rightarrow$ Water Surface \cite{nh2022study}. In most of cases,
NDWI values in range $[0.0, 0.2)$, indicating the humidity over the Saronicos gulf. Near the coastal areas the values are higher. The only exception
is observed in the image of 13th of September, 2017, where it is observed the abnormality in of the oil-spill in the middle of the Saronic Gulf.

\begin{equation}
\label{eq::NDWI}
    NDWI = \frac{NIR - SWIR-1}{NIR + SWIR-1}
\end{equation}

The WRI index is calculated by the Equation \ref{eq::WRI} and indicates the water ratio in range $[-1.0, 1.0]$ as follows:
$[-1.0, -0.75) \rightarrow$ Critical water areas; $[-0.75, -0.25) \rightarrow$ Normal water quality $[-0.25, 0.00) \rightarrow$ Wet ground / Vegetation; 
$[-0.25, 0.00) \rightarrow$ Wet ground / Vegetation; $[0.00, 1.00] \rightarrow$ Ground / Infrastructures. Based on the observations of the WRI in Figure 
\ref{fig::CaseStudyAnalysis::Results} is is clearly observed a periodic phenomenon of critical water areas among the Salamina coast line. Moreover, in the
image of 13th of September, 2017 critical values are observed in the area of the oil-spill (green box in Figure \ref{fig::CaseStudyAnalysis::Results}), as well 
as similar areas are indicated in the following weeks after the oil-spill. Another observation of the analysis, is that Landsat-8 products were unable to 
indicate the oil-spill due to low spatial resolution (i.e., 30m/px).

\begin{equation}
\label{eq::WRI}
    WRI-2 = \frac{GREEN - RED}{NIR + SWIR-2}
\end{equation}

The OSI (Oil-Spill Index) is referred in the work of \cite{Zakzouk2024novel} and is calculated by the Equation \ref{eq::OSI}. A first observation
is the difference of the products' values between Landsat-{8,9} and Sentinel-2 data. In Landsat-{8,9} products the index effectively distinguish 
the area of interest into water bodies ($values > 2.5$), vegetation regions ($1.9 < values < 2.5$), and building/infrastructures ($values < 1.9$).
In the Sentinel-2 products can be divided to the following ranges: $[0.75, 1.00) \rightarrow$ vegetation regions; 
$[1.00, 1.9) \rightarrow$ building/infrastructures; and $values > 1.9 \rightarrow$ water bodies. In the water bodies are additionally observed two trends:
(a) $[1.9, 2.5) \rightarrow$ indicate areas emitting strongly in blue spectral radiance; (b) $ > 2.5 \rightarrow$ indicate areas where water absorbs fully the 
blue spectral radiance. Considering the aforementioned observations in the product corresponding to the 13th of September, 2017, oil-spill patterns are
indicated.

\begin{equation}
\label{eq::OSI}
    OSI = \frac{RED + GREEN}{BLUE}
\end{equation}

Combining all the information that is observed by all these indexes, the critical area is bounding by 
the green box as illustrated in Figure \ref{fig::CaseStudyAnalysis::Results}. In addition, it is necessary
to mention that this time-series analysis indicates other information as well. For instance, among Salamina's
coast line is is observed a periodic and recursive pattern of dark pixel values in NDWI and WDI products. This
phenomenon indicates additional areas of interest that may be useful for specific use cases related to climate
change. Moreover, due to the repeatability of this phenomenon, it is safely to assume that is not correlated with
the oil-spill disaster, because it is also observed in images before the oil-spill.

\section{Discussion}
\label{sec::Discussion}
The main objective of this work is to introduce a new Data Cube implementation for coastal region monitoring named Ocean-DC.
The Ocean-DC framework handles several kinds of data combining both EO and non-EO products for generating a NetCDF file that contains
all the necessary information over a predetermined area of interest. In addition, during the cube's generation it calculates the most
common remote sensing products to help more the end-users by providing them with a ready to use product. As an example, in this paper
it was generated a 4D Ocean-DC for monitoring the oil-spill disaster that took place at 10th of September, 2017. The generated Ocean-DC
product successfully applied data from different sources (i.e., Landsat-{8,9} and Sentinel-2) to a common template and provided a strong
ready to use dataset for time series analysis. Thus, the advantages of the Ocean-DC framework are the following:

\begin{enumerate}
    \item It utilizes several EO and non-EO data for generating a ready to use product.
    \item It supports several EO data from different sources and harmonize/homogenize them to a common. This feature significantly accelerates
    the post-processing analysis by allowing the end-users to visualize several data using a single file.
    \item It calculates the state-of-the-art remote sensing data providing a ready to use product. This feature also accelerates the post-processing
    analysis by allowing the end-users to emphasize only on the interpretation of the generated products without spending time to calculate the products.
    \item The aforementioned characteristics allowing the usage of the Ocean-DC products by both experts and non-experts, because after the generation of
    the DC, the end-user have to visualize and interpreter the data.
    \item The outcome of the Ocean-DC framework is a NetCDF file, which is readable by the most common GIS software. Thus, it is supported by both
    free and open-source and commercial GIS software, allowing to end-users to use their preferred software.
\end{enumerate}

Summarizing the discussion, it is worth mentioning that the Ocean-DC framework is not finalized yet. This means, that the framework is updated frequently
by expanding this framework with additional features. Moreover, further testing is necessary for indicating further the efficiency of the Ocean-DC framework.
Thus, possible future works will combine the Ocean-DC framework with Machine Learning and Deep Learning algorithms for generating more products like
cloud and shadow masks, expanding more the applicability of the Ocean-DC products for coastal monitoring. Moreover, other application like in-land vegetation,
deforestation, forest fires, or flooding monitoring will be examined as well.

\section{Conclusions}
\label{sec::Conclusions}
In conclusion, this work introduces a new DC framework that handles several EO and non-EO data 
from different sources for periodic monitoring the environmental hazards and climate change effects
over the coastal areas. The framework may still be under development, however, it can be effectively
utilized for time-series analysis and the visualization of disasters over the coastal areas, as 
indicated in the presented case study. As a result, the Ocean-DC framework can effectively utilized in
monitoring phenomena applications. However, the Ocean-DC as an emerging and currently in development
implementation will be updated frequently to support more applications by utilizing novel and smart
technologies.

\section{Acknowledgments}
This work is supported by the European Union Funded project OCEANIDS 
”User-driven applications and tools for Climate-Informed Maritime Spatial 
Planning and integrated seascape management, towards a resilient \& inclusive 
Blue Economy”, under the EU HE research and innovation program under GA No. 101112919.

\bibliographystyle{unsrtnat}
\bibliography{main}

\end{document}